# Microwave Schottky diagnostic systems for the Fermilab Tevatron, Recycler, and CERN LHC

Ralph J. Pasquinelli, Fermi National Accelerator Laboratory[*], P.O. Box 500, Batavia, Illinois 60510, USA
Andreas Jansson, European Spallation Neutron Source, Lund, Sweden

A means for non-invasive measurement of transverse and longitudinal characteristics of bunched beams in synchrotrons has been developed based on high sensitivity slotted waveguide pickups. The pickups allow for bandwidths exceeding hundreds of MHz while maintaining good beam sensitivity characteristics. Wide bandwidth is essential to allow bunch-by-bunch measurements by means of a fast gate. The Schottky detector system is installed and successfully commissioned in the Fermilab Tevatron, Recycler and CERN LHC synchrotrons. Measurement capabilities include tune, chromaticity, and momentum spread of single or multiple beam bunches in any combination. With appropriate calibrations, emittance can also be measured by integrating the area under the incoherent tune sidebands.

## I. Introduction

The Schottky detector is based on the stochastic cooling pickups[1] that were developed for the Fermilab Antiproton Source Debuncher cooling upgrade which was completed in 2002. These slotted line waveguide pickups have the advantage of large aperture coupled with high beam coupling characteristics. For stochastic cooling, wide bandwidths are integral to cooling performance. The bandwidth of slotted waveguide pickups can be tailored by choosing the length of the pickup and slot spacing. The Debuncher project covered the 4-8 GHz band with eight bands of pickups, each with approximately 500 MHz of bandwidth. For use as a Schottky detector, bandwidths of 100-200 MHz are required for gating, resulting in higher transfer impedance than those used for stochastic cooling.

Bunched beam stochastic cooling was attempted in the Tevatron[2] in the mid 1990s but was unsuccessful due to the large coherent signals produced by the bunched beams. Observance of the Schottky signals at microwave frequencies revealed incoherent as well as coherent frequency bands containing relevant transverse and longitudinal characteristics of the beam. Depending on the bunch fill pattern, the coherent signals fluctuate significantly from revolution line to line, but in all cases, a high instantaneous dynamic range is required to maintain linear performance. Figure 1.

Many diagnostic systems for the measurement of beam parameters utilize resonant circuits to increase beam sensitivity. Such resonant characteristics do not allow the investigation of individual beam bunches spaced by tens of nanoseconds due to the ringing that results from the higher Q. If a diagnostic system could be made wide band so that gating would enable the extraction of bunch-by-bunch data and retain adequate signal to noise ratio, an investigation of individual or multiple bunch characteristics could be explored.

In 2003 after the completion of the Debuncher stochastic cooling project, a Schottky diagnostic system for the Fermilab Tevatron and Recycler was proposed and implemented.[3] The center frequency chosen was 1.7 GHz based on the required minimum beam aperture of the pickup (70mm x 70 mm) and the availability of surplus microwave hardware from the recently decommissioned 1-2 GHz Stacktail cooling system in the Accumulator. With successful operating systems at Fermilab[4, 5], it was suggested that a similar system would be beneficial to the CERN LHC, which was under construction at the time. In 2004, a proposal was made to the LARP[6] collaboration that such a system be developed under their auspices. The LHC system would be centered at 4.8 GHz, the 12th harmonic of the LHC RF system. At this frequency, upper and lower sideband Schottky signals do not overlap, the pickups satisfy the minimal beam aperture requirement of 60x60 millimeters, and the vacuum vessels would fit between the 194 millimeters beam center-to-center spacing. The LHC system[7] was installed in 2007 and successfully commissioned in the first physics run in 2010.

## II. SCHOTTKY SIGNALS

In a machine with longitudinally focusing RF, parameterized by the synchrotron frequency $\Omega_s$, the signal from a single particle in a transverse position pick-up can be expressed as[8]

$$d_i(t) = e\, f_0 \sum_{n=0}^{\infty} \sum_{p=-\infty}^{\infty} \hat{a}_i \,\mathrm{Re}\left\{ \mathrm{J}_p\left((n \pm q)\omega_0 \hat{\tau}_i\right) \exp\left( j\left[ \left((n \pm q)\omega_0 + p\,\Omega_s\right)t + p\psi_i + \varphi_i \right] \right) \right\}$$

where $n$ is the revolution harmonic and $p$ is the synchrotron sideband index. The sum over upper and lower sidebands as indicated by $n\pm q$ is implied. Here, $\hat{a}_i$ and $\varphi_i$ denote the initial transverse amplitude and phase, $\hat{\tau}_i$ and $\psi_i$ is the initial longitudinal amplitude and phase, $q$ denotes the fractional machine tune, $e$ is the particle charge, $\mathrm{J}_p$ the Bessel coefficient and $f_0$ ($\omega_0$) the revolution frequency. The formula gives the beam dipole moment, which combined with the transfer impedance of the pick-up gives the output voltage. The expression describes two betatron sidebands (around $(n+q)\omega_0$ and $(n-q)\omega_0$), each consisting of an infinite number of synchrotron satellites. The betatron sidebands come from the transverse oscillations of the particle (AM modulation), while the split into the synchrotron satellites arises from the variations in revolution time of the particle due to the longitudinal oscillations (FM modulation).

A stable beam (i.e. one where the beam distribution is not changing from turn to turn) by definition has a uniform distribution in the phases. In this case, the average (macroscopic) signal vanishes. If one considers the rms signal power at a given frequency, contributions from different particles add incoherently due to the random phase. This is referred to as the Schottky signal.

It can be shown from the above equation that the total signal power in the two-betatron sidebands is the same and proportional to the rms transverse amplitude $\langle \hat{a}_i^2 \rangle$, which in turn is a measure of the transverse emittance. Due to the (anti) symmetry of the Bessel function, the betatron sidebands are symmetric and by determining their center, one can measure the fractional tune. Moreover, with the knowledge that $J_p(x)\sim 0$ for $|p|>x$ if $x$ is large, it can be shown that the effective half width of the betatron sidebands produced by a single particle is given by $(n \pm q)\omega_0 \hat{\tau}_i \Omega_s$, where $\hat{\omega}_i = \omega_0 \hat{\tau}_i \Omega_s$ is simply the amplitude of the revolution frequency variation for the particle. Keeping in mind that $\Delta f_i / f = \eta \, \Delta p_i / p$, and $\Delta q_i = Q\xi \, \Delta p_i / p$, the rms width of the betatron sidebands for the beam is given by

$$\Delta f_{\beta\pm} = f_0 \frac{\Delta p}{p}\left[(n \pm q)\eta \pm \xi Q\right].$$

if chromaticity is the only source of tune spread in the beam. Hence, the momentum spread can be determined from the average width of the sidebands and the chromaticity from the difference in upper and lower sideband width.

The above formula indicates that the synchrotron satellites are infinitely narrow. In reality, there is a small spread in synchrotron frequencies in the beam. The width of synchrotron lines then increase with the line index $p$. This may cause synchrotron satellites to overlap for large momentum spread.

## III. PICKUP DESIGN CONSIDERATIONS

The pickup is essentially a sandwich of three waveguides separated by coupling slots in a thin metal foil, figure 2. The center waveguide is that in which the beam circulates. The output of the outer two waveguides when added 180 degrees out of phase with the use of an external microwave hybrid circuit provides the difference mode containing transverse characteristics of the beam. The pickup is bidirectional, which allows it to be used for both protons and antiprotons in the Tevatron. The hybrid output sum mode signal is also utilized for a beam present witness signal for the gating system. Bandwidth selection of the individual systems was based on the bunch spacing. In the Fermilab Recycler, beam is bunched by barrier buckets, which do not present a significant bandwidth restriction. The Tevatron bunches are in buckets that are18.8 nanoseconds in length and separated by 21 buckets, or 396 ns. The LHC has an ultimate bunch spacing of 25 ns.

To minimize the types of pickups to be designed and fabricated, both the Recycler and Tevatron have the identical pickups centered at 1.7 GHz. A 3 dB bandwidth of 100 MHz was chosen to provide adequate response for the bunch-by-bunch gating requirement in the Tevatron. Figures 3 and 4 show the calculated and measured pickup performance. In the Tevatron, both protons and antiprotons circulate on separated helical orbits through the same pickup. While the pickup was measured to have a directivity of 12 dB (using forward antiprotons and reverse protons in the Recycler), it was imperative to locate the pickups in a region of the Tevatron where protons and antiprotons do not intersect. Preferably in a spot where crossings are tens of nanoseconds apart, easing the gating speed requirements. Such a location would allow the gating system to cleanly separate the two beam signals. In the Tevatron, proton bunches contain three to six times more particles than the antiproton bunches. The E17 straight section of the Tevatron had available space for two vacuum vessels (one horizontal, one vertical) and meets the beam crossing separation criterion. Identical vacuum vessels were installed in the Recycler 61 sector.

The large revolution signal forces the signal processing electronics to maintain a high instantaneous dynamic range, avoiding signal compression in active components. The best common mode rejection occurs for beam that traverses the pickup on the electrical center orbit. Unfortunately, in the Tevatron with helical orbits, both beams cannot be on this same orbit, resulting in larger common mode for any off axis beam. The pickups are motorized to allow adjustment for maximum dynamic aperture, but they are not adjusted for best common mode rejection of either beam.

For the LHC, the minimum aperture specification is 60x60 millimeters. A further restriction was the pickup was not to be moved physically by means of remotely controlled motorized stands. This last requirement is an attempt to minimize, the number of movable devices in the LHC that could accidentally intercept the beam with a potential 400 mega Joules of stored (beam) energy. The LHC has two separate beam pipes; hence, location was not as critical as in the Tevatron as there is no bunch crossing restrictions. Four pickups were constructed, two in each beam located near access point 4. A remotely controlled continuously variable attenuator and delay line are installed in one of the pickup outputs. The best common mode rejection is tuned empirically on the beam signal by adjustment of these devices.

The Tevatron/Recycler pickups utilize microwave absorber at both ends. These were added to reduce the propagation of waveguide modes generated by the beam. The material is TT2-111R from Trans-Tech Inc. In addition to microwave absorption, this material is compatible with the ultra high vacuum requirements for a storage synchrotron. During the design phase of the LHC pickups, calculations revealed a significant beam wake field heating of such absorbers (with the anticipated beam intensities) could result in poor vacuum performance. The lack of available cooling water at point 4 and associated plumbing complications to the pickup vessel design required dropping the use of absorbers for the LHC systems. No deleterious effects have been observed with 400 bunches of $10^{11}$ protons per bunch (November 2010). This is approximately one fifth of the expected beam intensity.

## IV. SIGNAL PROCESSING

Early bunched beam-cooling measurements displayed large coherent signals at harmonics of the revolution frequency at microwave frequencies. A simple Fourier transform of the beam envelope would indicate that revolution lines should not be present at frequencies above 1 GHz. This turns out not to be the case as intra bunch instabilities[9] generate coherent lines observed in the Tevatron at FNAL, SPS at CERN, RHIC at Brookhaven, and HERA at DESY in Hamburg. To reduce any nonlinearities and intermodulation distortion, the processing electronics must have large instantaneous dynamic range to cope with these coherent harmonics.

All of the relevant characteristics of the beams are contained in every revolution harmonic. The final baseband bandwidth therefore does not need to exceed the width between two revolution lines, which for the Recycler, Tevatron, and LHC is 89 KHz, 47 KHz, and 11 KHz respectively. The gating requirement, though, requires significant bandwidth to allow individual bunch gating capability. The Fermilab systems have a pickup bandwidth of 100 MHz. The tighter 25-nanosecond bunch spacing in the LHC required a pickup with 200 MHz minimum bandwidth. (This has yet to be measured at CERN, as it requires a remotely controlled spectrum analyzer in the tunnel. The 4.8 GHz pickup signal would experience significant loss on the cable between the tunnel and surface service building.) Once gating is accomplished, the need for wideband signal processing is no longer essential. Band limiting cavity and crystal filters render a reduced bandwidth thus improving noise performance and limiting the integrated power from multiple revolution bands.

The pickup preamplifier should have the lowest available noise figure. However, more importantly, a very high output power compression rating to cope with the large dynamic range. These two characteristics are not normally found simultaneously in commercially available amplifiers and a compromise is made in favor of compression levels. For the Tevatron and LHC, these first amplifiers have a +25 to +27 dBm 1 dB compression level and noise figures ranging between 2 and 4 dB.

The signal processing hardware designed for the Tevatron and Recycler originally utilized single sideband (SSB) down conversion employing ninety-degree hybrids. The SSB method was essential for retaining chromaticity information (double sideband conversion causes upper and lower sidebands to overlap at baseband) and was in use for several years in both accelerators. The SSB technique provided adequate performance for both Fermilab accelerators, but does have a limited dynamic range of 60-70 dB. During the LHC Schottky system design phase, it was desired to create a system with the highest possible instantaneous dynamic range. It was feared the coherent signal in the LHC could be significantly worse than that observed in the Tevatron. In keeping with the requirement to retain chromaticity measurements as part of the diagnostic, the original SSB technique was abandoned in favor of a narrow band crystal filter single side band system as suggested by CERN colleagues. Figure 5 is a simplified block diagram of the triple down conversion electronics. The crystal filtering method has improved dynamic range over the SSB and when tested in the Tevatron yielded improved performance. The Tevatron was retrofitted with the crystal filter electronics and became the test bed for the LHC hardware.

Each of the revolution lines contains details of beam parameters that are easily resolved in the frequency domain. Spectral analysis can be performed directly on the microwave signal or at lower frequencies with down conversion utilizing off the shelf spectrum analyzers. Spectrum analyzers are typically designed for wide frequency ranges and have limited instantaneous dynamic range. Price for these analyzers is commensurate with performance and ranges from thousands to tens of thousands of dollars per channel. For the Schottky system, spectral analysis is performed at baseband. The custom down conversion electronics are optimized for this narrow band performance with 100 dB of instantaneous dynamic range, figure 6.

Whether the signal is down converted with commercial or custom hardware, front end gating is essential for bunch-by-bunch signal acquisition. All of the components in the signal processing electronics are commercially available. The fast gate consists of a Mini Circuits ZASWA-2-50DR switch[10]. This device has an analog bandwidth of DC-5 GHZ, which satisfies both FNAL and CERN Schottky system frequencies, a turn on turn off rise time of 5 ns, and 50 dB minimum dynamic on to off isolation. Once gating is complete, reduction of signal bandwidth is possible to reduce noise and peak integrated signal power.

Conventional high-level double balanced mixers are used for each stage of down conversion. The high level mixer is required due to large dynamic range of the Schottky signal. More importantly is the selection of local oscillators (LO) for their phase noise performance. The Schottky signals are typically 10 dB or less above the system noise floor. The machine revolution frequencies range from 89 KHz (Recycler), 47 KHz (Tevatron) to 11 KHz (LHC) indicating the betatron sidebands will be closely spaced in frequency to the revolution lines. Figure 7 shows a typical IF at 400 MHz in the Tevatron utilizing two different sources for the first LO. The poor phase noise due to the LO in the green trace virtually obscures the Schottky signal. The phase noise on the blue trace clearly demonstrates the need for clean LOs as they establish the minimum detectable signal. Phase noise of oscillators is worse closer to the fundamental frequency output. With the lower revolution frequency at the LHC, sidebands are within a few KHz of the revolution lines forcing the use of premium performance LOs at each stage of down conversion. Most recent tunes in the Recycler, Tevatron, and LHC are of order 0.465, 0.585 and 0.320 respectively.

The final stage of down conversion employs 15 KHz crystal filters, two in series for maximum out of band rejection. For the LHC this bandwidth will pass at least one revolution line. In the Tevatron, LO frequencies are chosen to center the two-betatron sidebands in the pass band and no revolution lines are present. Fermilab chose to use two Agilent dual channel Vector Signal Analyzers (VSA) to perform the fast Fourier transform (FFT) of the base band data, one each for antiproton and proton signals, figure 8. The LHC had already developed a baseband digitizer card with adequate dynamic range and frequency resolution for other instrumentation; hence, this hardware was selected for the FFT at CERN.

An extensive system noise analysis was performed for the signal processing chain. A 100 dB instantaneous dynamic range was achieved by the careful selection of components, matching, and electromagnetic shielding. A calibration signal is also injected into the processing electronics as a diagnostic aid for monitoring gain variations over time. The Tevatron system utilizes both up and down stream pickup outputs for proton and antiproton signals. The Tevatron calibration signal is injected before the first preamplifier. In the LHC, only one output from the pickup is used, leaving the other available for calibration signal injection. This has the added benefit that the pickup response and electronics transfer function can be measured in the calibration mode. Figure 9 depicts the LHC calibration schematic.

## V. RECYCLER PERFORMANCE

The coherent nature of bunched beams complicates the measurement of the pickup's beam sensitivity. In the early stages of commissioning the Recycler prior to the availability of antiprotons, it was possible to inject reverse protons. Due to the bi-directionality of the pickups, the transfer function could easily be measured on debunched coasting beam. (Refer to figure 4)

The original SSB hardware remains in the Recycler. An Agilent vector signal analyzer (VSA) is utilized to provide the fast Fourier transform (FFT) of the base band signal. In the Recycler, the Schottky diagnostic is the principal measuring tool for tunes and momentum spread. With a single species of Recycler beam, i.e. antiprotons, it is possible to dramatically reduce the common mode signal by centering the pickup on the beam. A power meter is employed to minimize pickup power across the band. Some 3 to 6 dB total power reduction is typical. Figure 10 illustrates significant reduction in the revolution line with a centered tank. This capability in the Recycler reduces the need for the high dynamic range required in the Tevatron and LHC.

At 1.7 GHz, the Schottky system operates in the same band as the longitudinal stochastic cooling system. At selected intervals when the Schottky tune is measured, the cooling system is gated off to reduce signal suppression cross talk between the cooling and Schottky systems. The Recycler tune is close to the half integer. Schottky bands begin to overlap at 1.7 GHz with beam currents exceeding $1x10^{12}$. The resolution bandwidth of the FFT must be set sufficiently narrow (<3KHz) to resolve the two sidebands. Measurements of both planes take about a minute with the existing controls application.

## VI. TEVATRON PERFORMANCE

The Tevatron Schottky system has been in continuous operation since 2003. Before the commissioning of this instrument, there was no means of measuring the antiproton tunes in the Tevatron. Co-rotating beams of protons and antiprotons in the same beam pipe made for difficult separation of the two signals. The high bandwidth of the Schottky system along with location of the pickups away from crossing points allowed a gating system to separate the signals. The primary function is measurement of tunes at collisions through the store. This is accomplished with an open access client (OAC) application that continually monitors and logs tunes, chromaticity, and momentum spread. An emittance value is also produced, but is not used since there are more direct methods to measure emittance in the Tevatron. Most notably, the beam-beam antiproton tune shift is adjusted based on the tune measurement. Figure 11 shows the variation in tune before and after using the Schottky tune measurement for antiproton tune compensation over a one-month period. The monitor guides the operations staff in manual adjustments of tune, keeping tune variation to a 0.005 window shown in the second half of the plot.

The individual bunch gating capability of the Schottky system has been utilized in studies with the electron lens. The electron lens has been used to alter the tunes of individual bunches. Measurement of the induced tune shift is accomplished with the Schottky monitor. The Schottky monitor has also been used to study the tune variation between different bunches in the three 12 bunch trains of a typical store. Figure 12 and 13 show significant tune and chromaticity variation for leading and trailing bunches.[11] In the Tevatron, proton bunch intensities of $3x10^{11}$ and antiproton intensity of $8x10^{10}$ are nominal. With averaging, single bunch protons and two or more bunches of antiprotons are easily resolved.

Due to the width of the betatron bands at 1.7 GHz, it is not possible to resolve the two normal mode frequencies. Rather, the devices rely on geometry to resolve the horizontal and vertical motion, and the tunes are derived from the center of the frequency distribution in each plane. However, in a coupled machine operating close to the coupling resonance, the normal modes may be inclined. This causes a mixing of the signal from the two normal modes into the horizontal and vertical signals of the Schottky. This effect tends to bring the observed tunes closer together than the normal mode tunes, counteracting the normal tune-separating effect of coupling. It can be shown that in a simple smooth approximation, the two effects cancel exactly[12]. In the Tevatron, the cancellation is only approximate, but for practical purposes, it can be said that the 1.7 GHz essentially measures the uncoupled tunes of the machine.

The average tune spread reported by the Schottky OAC agrees well with the bunch length based momentum-spread measurement for protons, but for antiprotons, a small offset has been observed early in stores, figure 14. This can be attributed to the additional tune spread arising for beam-beam, since it scales as the beam-beam parameter and has about the expected magnitude.

In the Tevatron, for the first few minutes after beams are set into collisions, the Schottky beam signals appear to be "boiling" furiously as seen in both time and frequency domains and stable tune measurements have not been reliable. A number of studies and hardware modifications to the system have been attempted to alleviate this condition, but it does not appear to be due to signal compression or nonlinearities in the processing electronics. The first LO in the system is designed to track the RF up the energy ramp in hopes that tunes could be measured. This LO is a custom device from Miteq Inc., which locks to the Tevatron RF, but does not have the best phase noise performance. This same boiling occurs during the energy ramp and tunes appear and disappear randomly. Better performance at energy flattop could be obtained with a stand-alone signal generator with improved phase noise.

## VII. LHC PERFORMANCE

Initial commissioning of the LHC Schottky system was completed in October of 2010 and is now considered an operational instrument for CERN's proton-proton/lead-lead collider. Tune and chromaticity measurements for both protons and lead ions has been demonstrated and verified against other instruments measuring these same parameters. Figures 15 and 16 show typical results for protons and lead ions. Note that with protons, the coherent line is sometimes present on the betatron sidebands. The resolution of the CERN system indicates that these lines are synchrotron satellites. These sideband satellites have not been observed in the Tevatron. In contrast with the Tevatron, the LHC Schottky system has also proven to be a valuable measurement tool during the beam energy ramp.

The measurement of chromaticity is particularly useful at injection energy, where the decay of the higher mode dipole component translates into a chromaticity drift, which, if left uncorrected, can lead to negative chromaticity and instability, figure 17. The Schottky monitor system currently is also the only monitor capable of bunch-by-bunch tune measurements and was used extensively during studies of the electron cloud phenomena observed when testing 50ns bunch spaced operation towards the end of the 2010 run.

Lead ion signals were significantly cleaner than proton signals and seemed to suffer less from coherent contributions to the transverse spectra, allowing an on-line measurement of tune and chromaticity throughout the accelerator cycle. The proton spectra suffered from the added complication of the RF longitudinal blow-up. The beam blow up being performed purposely to maintain a relatively long bunch length throughout the ramp. This produced a wide longitudinal component at the revolution line, which completely swamped the transverse Schottky bands. As has been observed at the Tevatron, the cleanest spectra were obtained several tens of minutes after initiating collisions for protons.

An added advantage of the LHC Schottky monitor, with its high detection frequency at 4.8 GHz, is its immunity to the effects of the transverse damper. The transverse damper system is being used at a relatively high gain to combat instabilities and emittance blow-up. The damper induced interference to the standard tune system is not observed with the Schottky system. The LHC Schottky system is therefore being considered as a serious alternative to the standard tune system for providing data to auto tune feedback. Implementation for tune feedback will require a faster tune acquisition than is currently possible with existing data acquisition hardware plus a robust measurement throughout the ramp, even in the presence of longitudinal blow-up.

As of July 2011, commissioning of the Schottky system at the LHC continues with the CERN instrumentation specialists. There is much more to be learned with the Schottky diagnostics at LHC.

## VIII. CONCLUSIONS

All three Schottky systems are in continuous use at Fermilab and early commissioning at CERN. The Schottky system in the Recycler is the main mechanism for tune and momentum spread measurements. The ability to adjust for beam-beam tune shifts in the Tevatron has yielded improved beam lifetimes and integrated luminosity. At CERN, the Schottky system is being considered as a prime measurement tool for tune and chromaticity.

## ACKNOWLEDGMENTS

The authors wish to acknowledge the talents and support provided by the Accelerator Division at Fermilab and the AB Instrumentation Department at CERN.

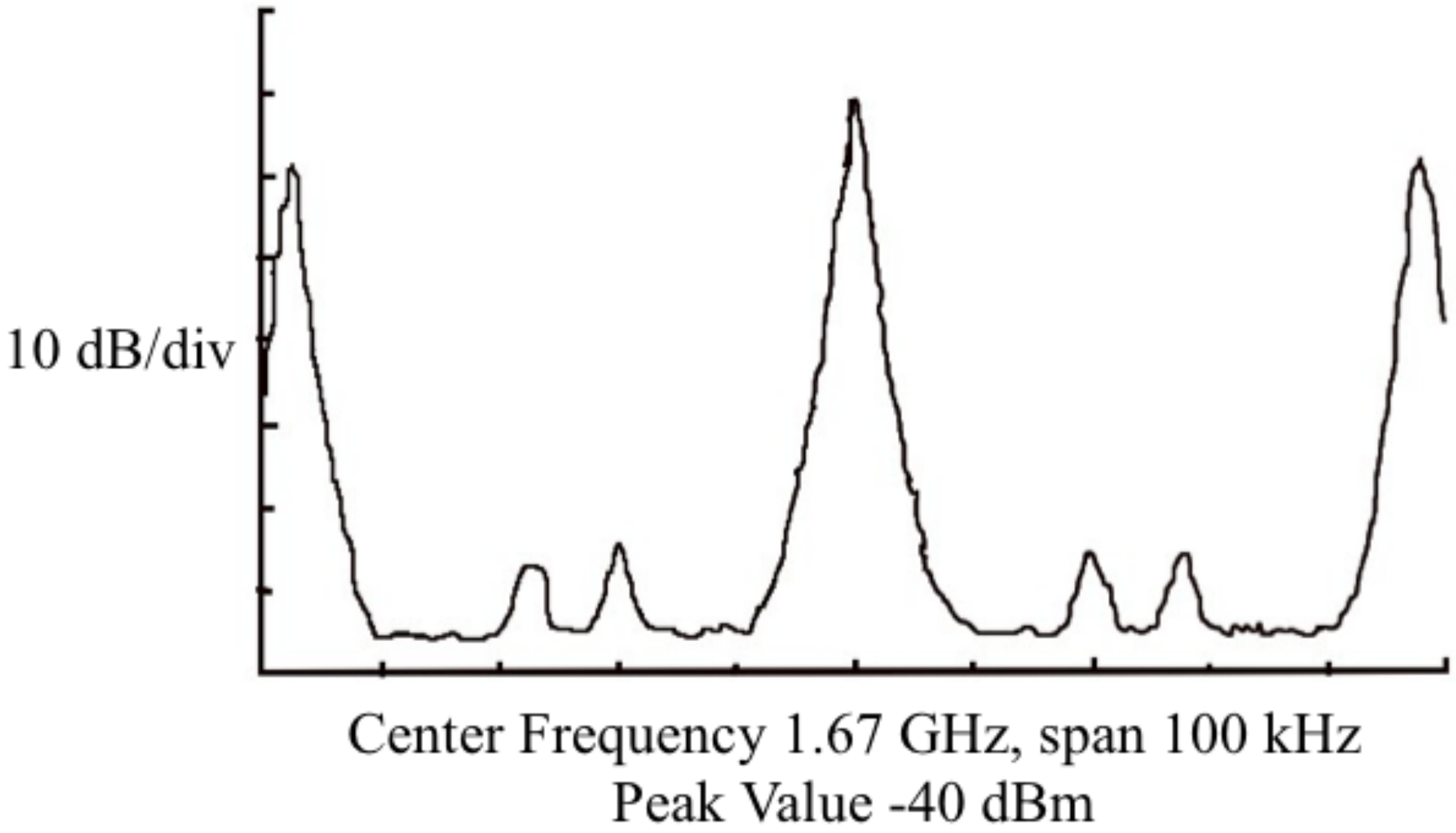

Figure 1. Tevatron Schottky pickup signal depicting large longitudinal coherent to betatron sideband ratio. 60-70 dB of instantaneous dynamic range is required to detect the betatron signals.

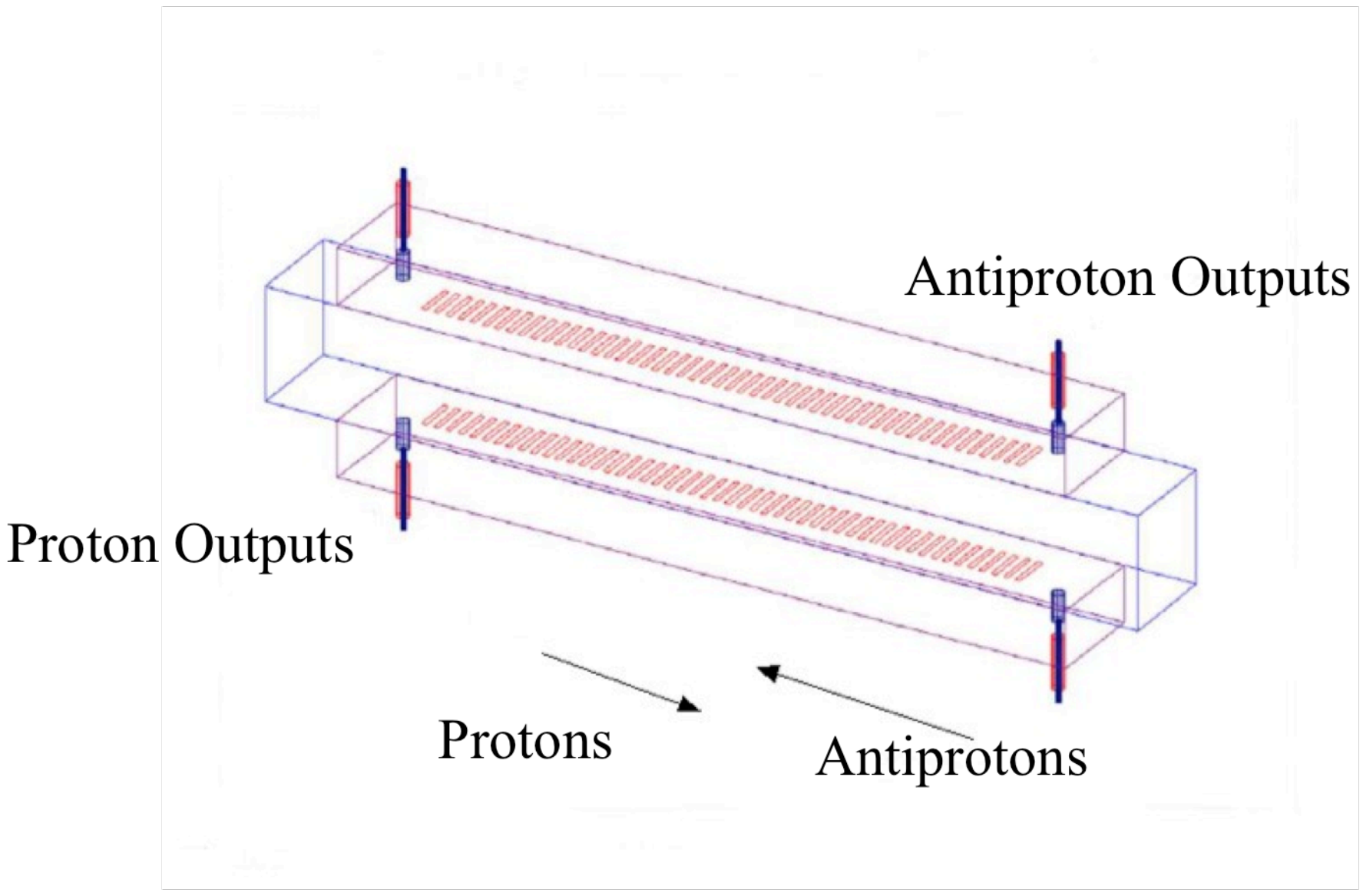

Figure 2. Slotted waveguide structure. Outputs can be used for bidirectional signals or injection of a calibration signal.

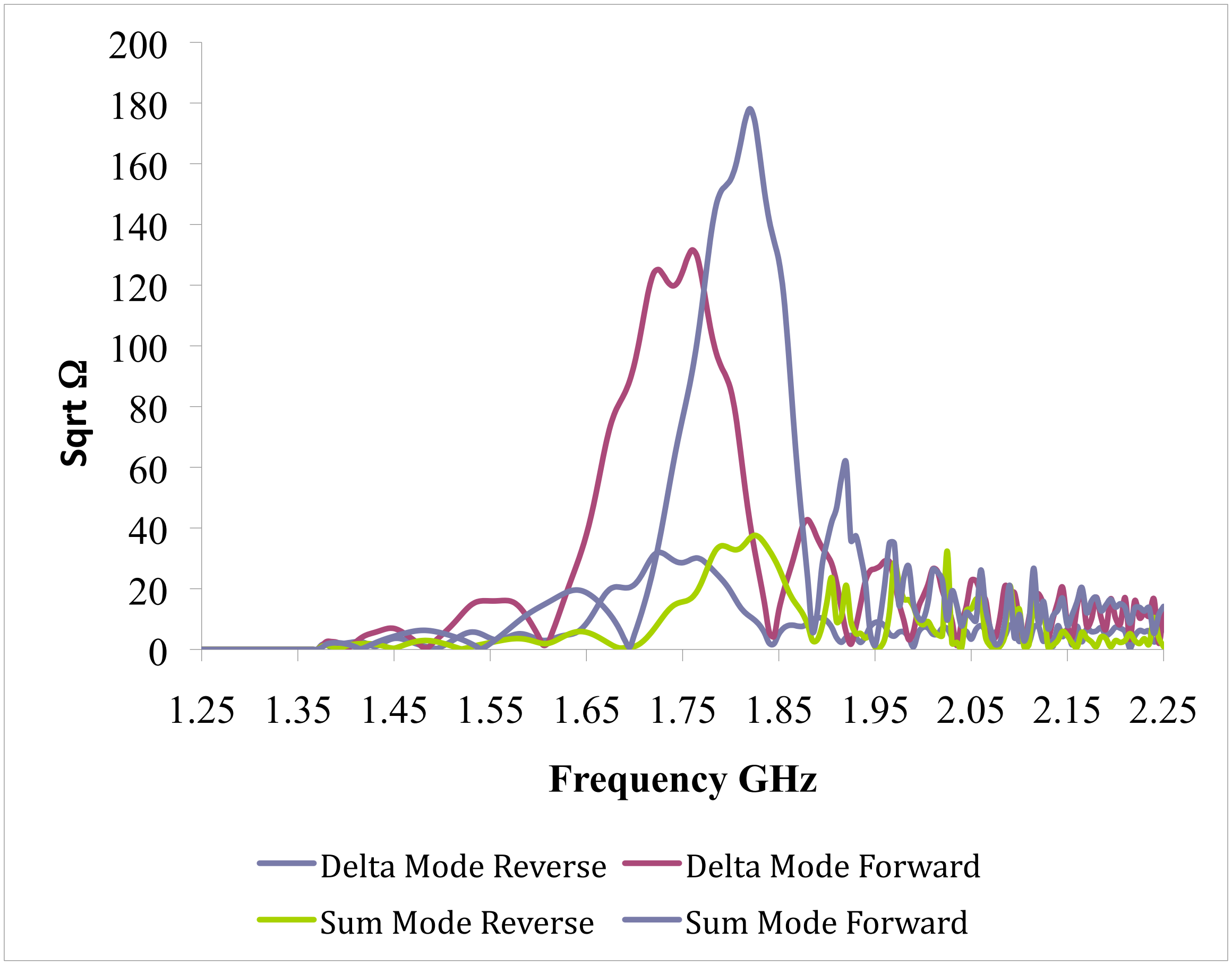

Figure 3. Calculated sum and difference response for slotted waveguide pickup for 1.7 GHz Schottky. Units are in square root Ohms to facilitate comparison with actual power measurements from a spectrum analyzer.

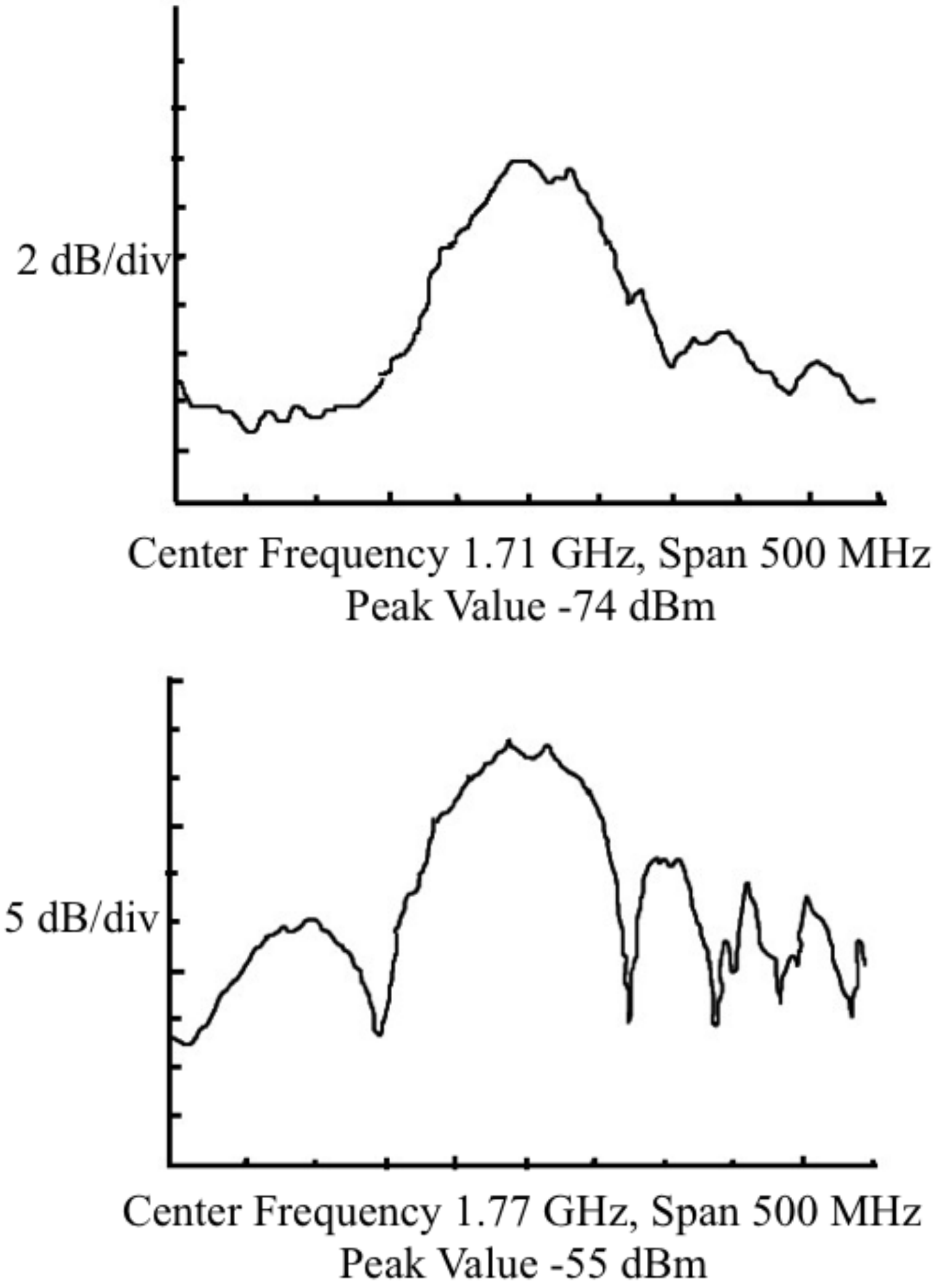

Figure 4. Spectral response of 1.7 GHz Schottky pickup for $1.4 \times 10^{11}$ DC coasting proton beam in the Recycler. Top difference mode, bottom sum mode.

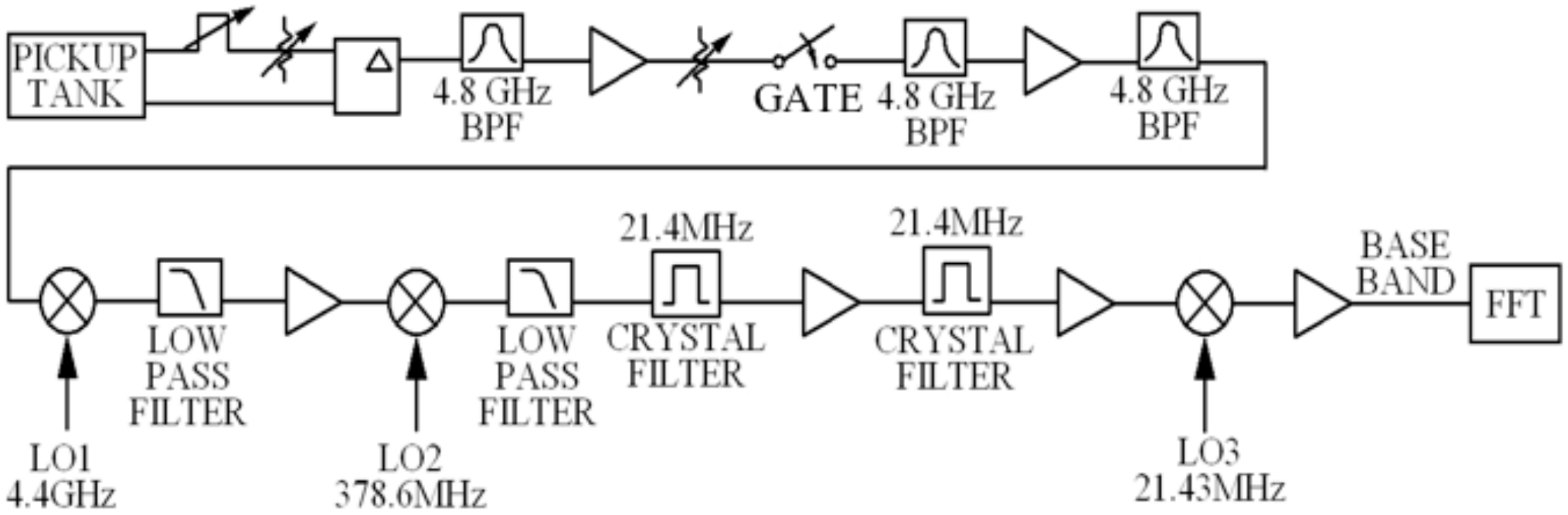

Figure 5: Block diagram of the triple down conversion signal processing system

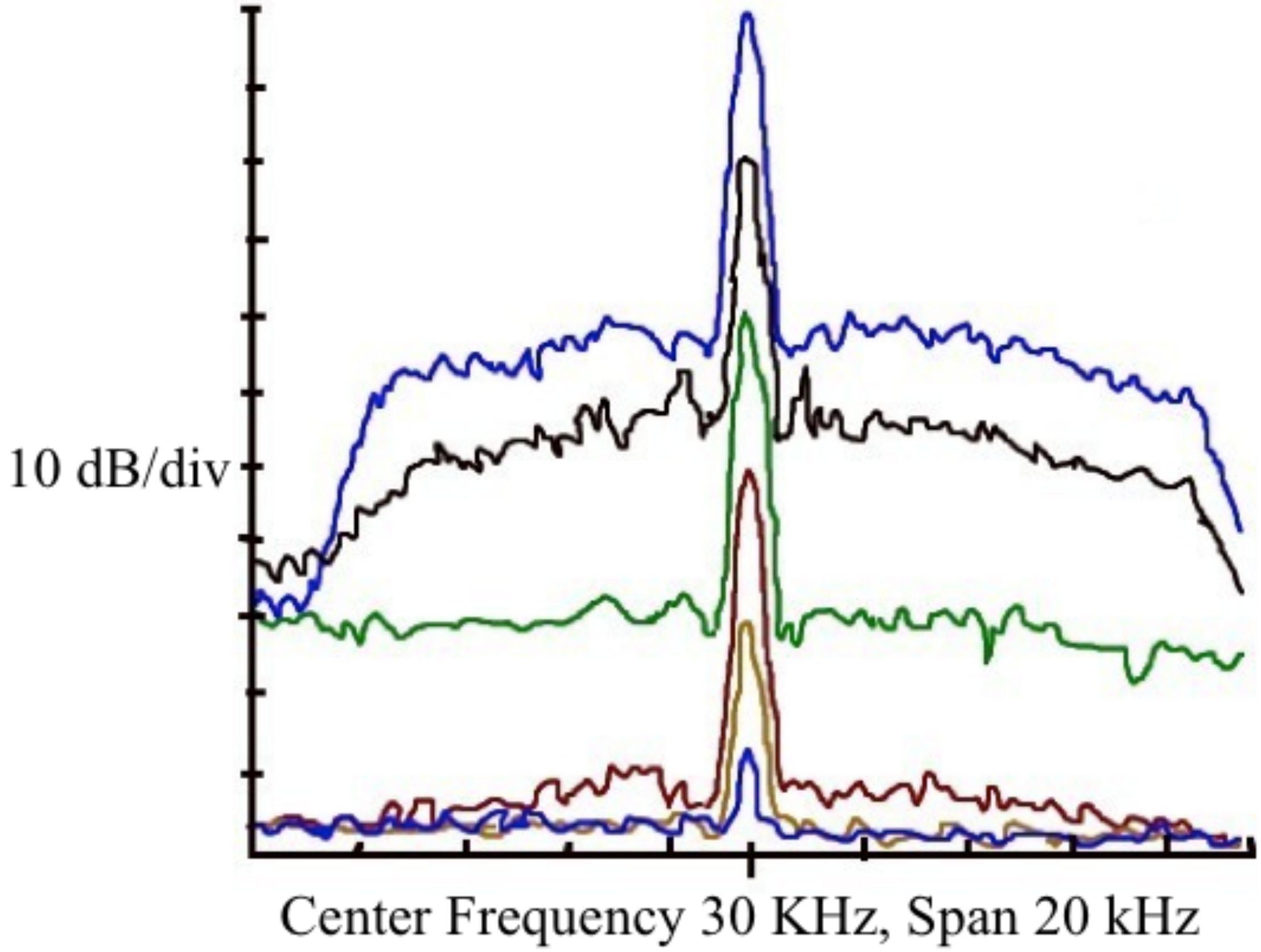

Figure 6. Measured 100 dB instantaneous dynamic range at baseband in Tevatron signal processing electronics utilizing triple down conversion. Final base bandwidth is limited by 15 KHz crystal filters. Input signal ranges from +10 dBm to -90 dBm in 20 dB steps.

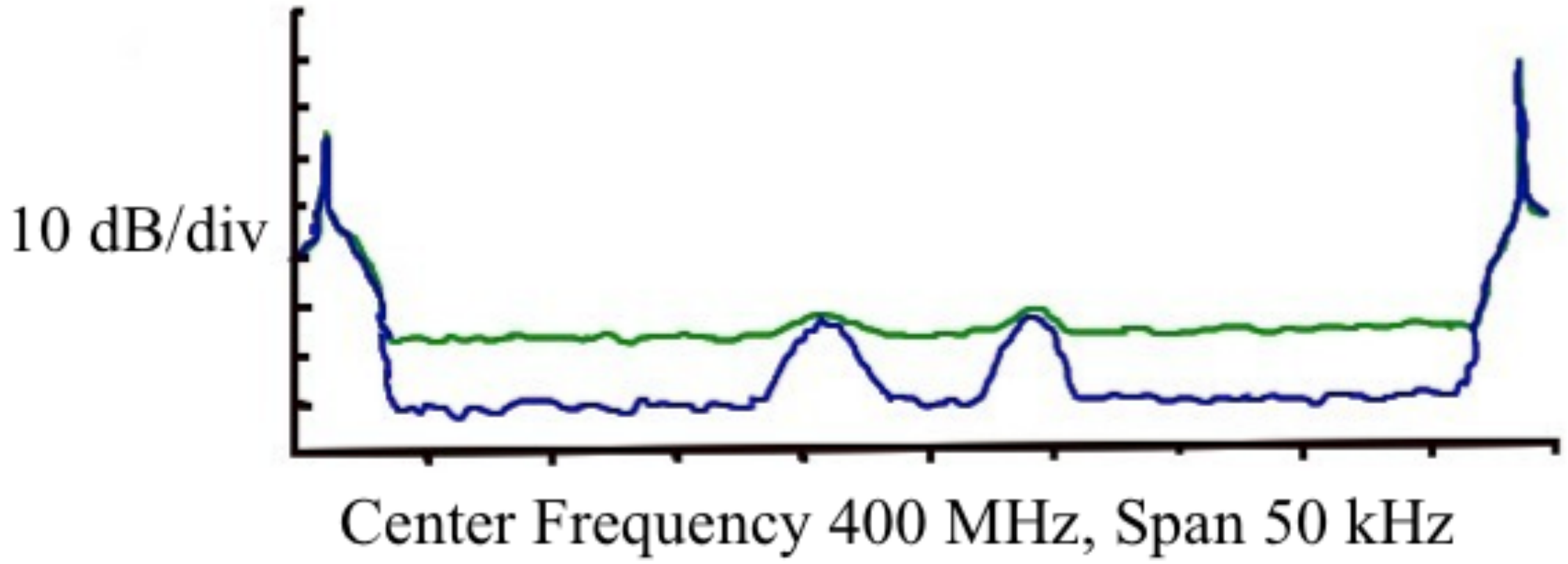

Figure 7. Effect of local oscillator phase noise on system sensitivity. 400 MHz IF sample front panel test point of Tevatron Schottky. Green HP8341A as first LO, Blue IFR 2042 as first LO.

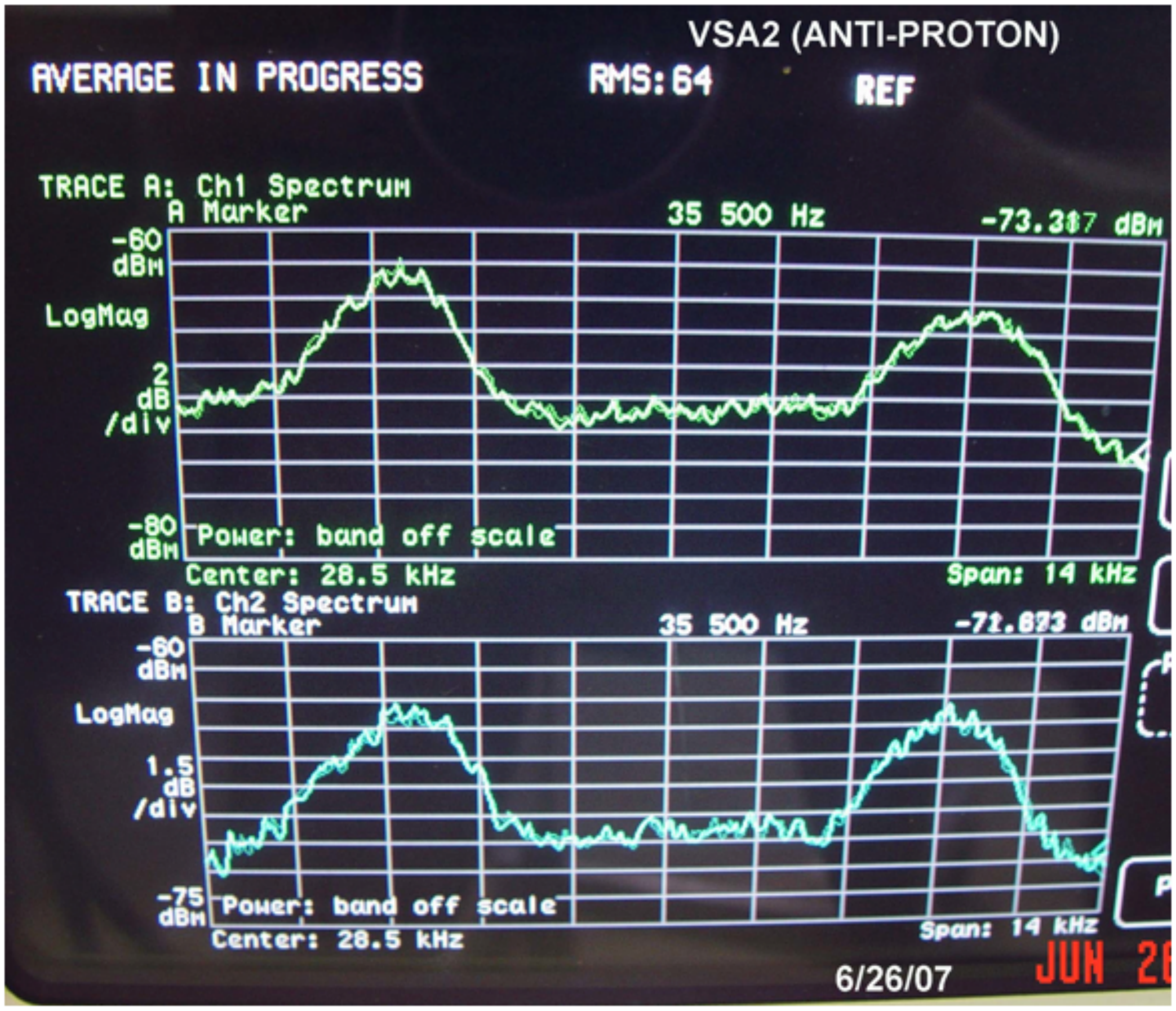

Figure 8. Tevatron antiproton Schottky baseband signal as displayed on Agilent VSA.

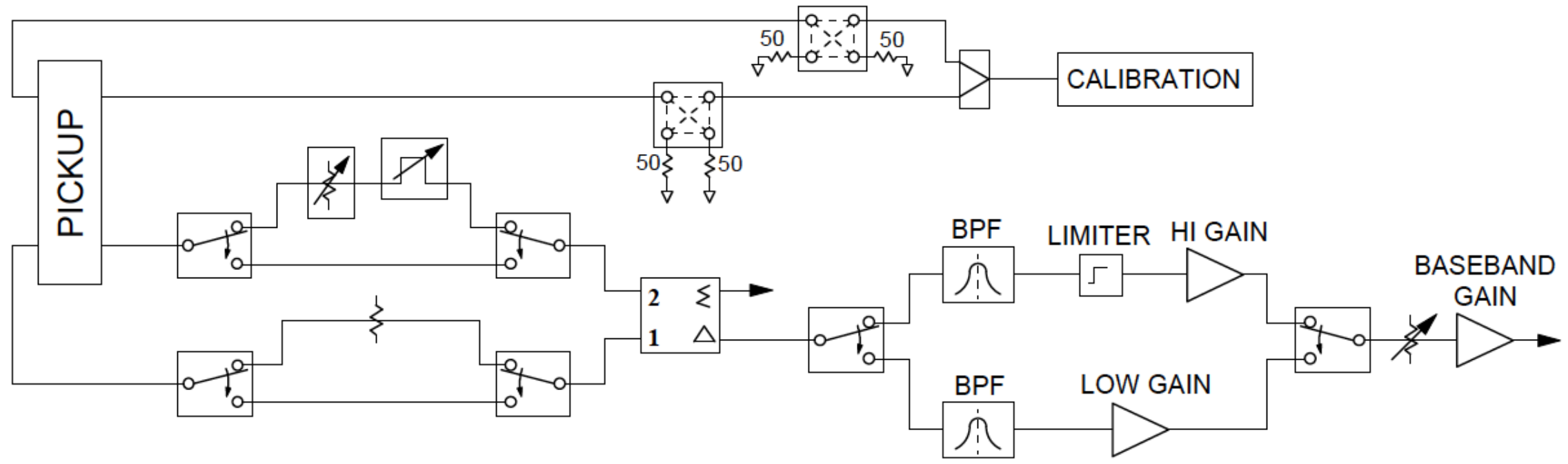

Figure 9. Calibration for LHC signal processing electronics. Pickup electronics has two processing paths with different gains and bandwidths to avoid signal compression for a variety of bunch intensities. Common mode rejection is improved by adjusting attenuator and delay in one leg of the pickup output.

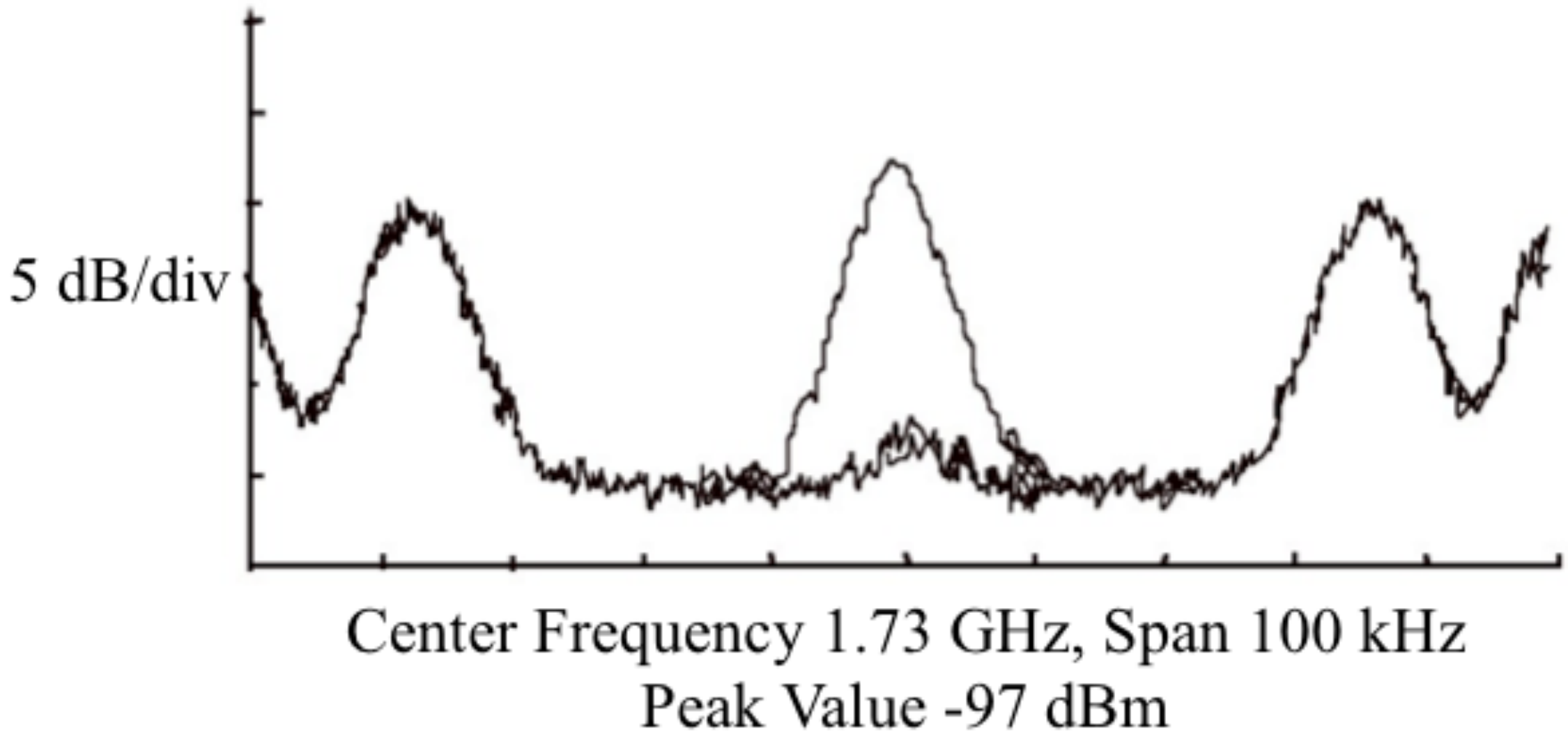

Figure 10. Single revolution harmonic of Recycler Schottky before and after pickup centering. In practice, pickup is centered by minimizing integrated pickup power,

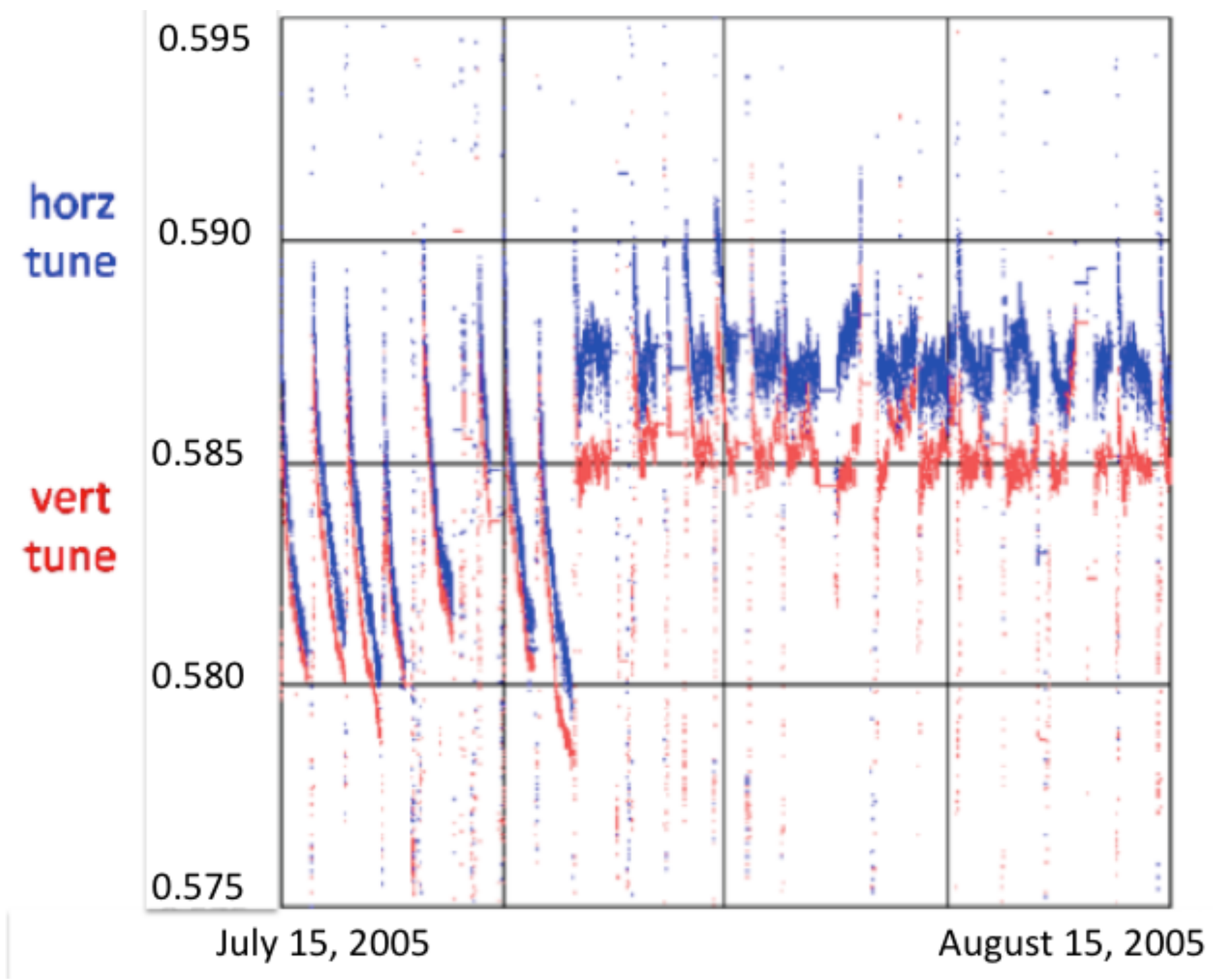

Figure 11. Tevatron antiproton beam tune shift. When the 1.7 GHz Schottky monitor was used beginning July 2005 to correct beam-beam tune shift, the tune excursions were minimized by manual tuning resulting in increased beam lifetime.

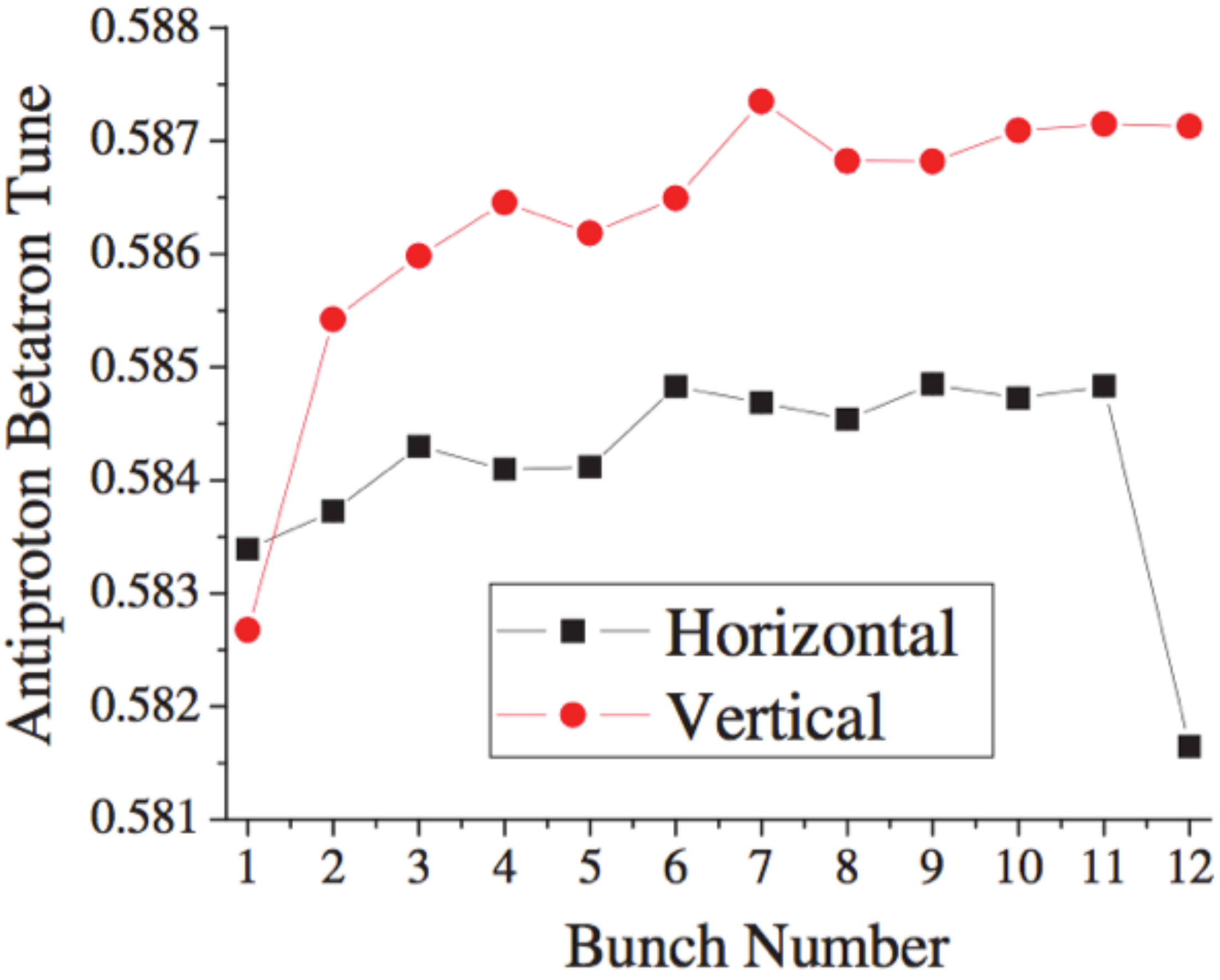

Figure 12. Horizontal and vertical antiproton tunes versus bunch number within a 12 bunch train measured by 1.7 GHz Schottky monitor at T=3 hours into store #3678 (July 27, 2004). The data was taken over a period of three hours starting three hours into the store. Three bunches are averaged in the same sequence of each train.

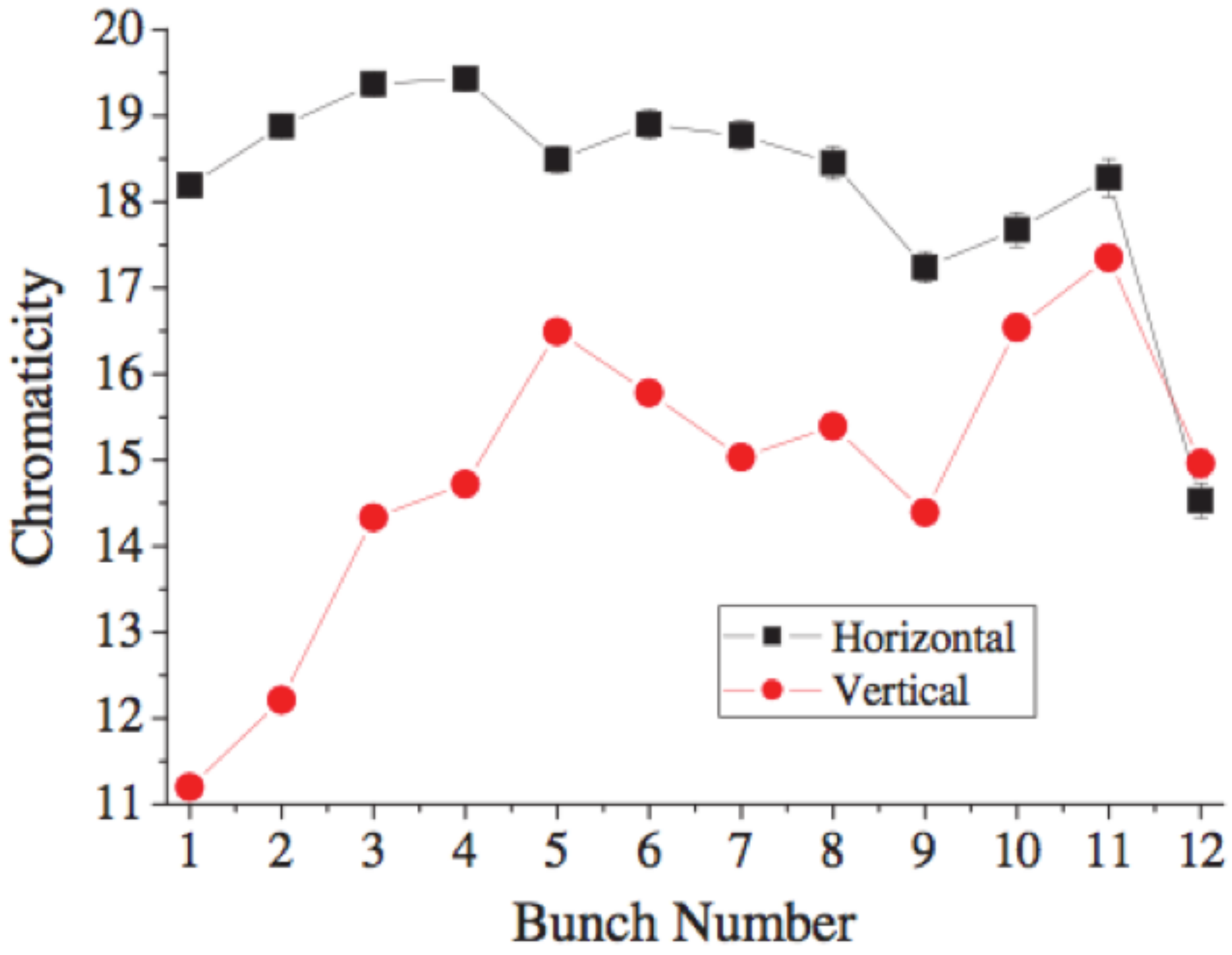

Figure 13. Antiproton chromaticities measured by the 1.7 GHz Schottky monitor versus bunch number for store #3678 (July 27-28, 2004). The chromaticities were assumed to be constant, and so the measurements were averaged over the entire store. Three bunches are averaged in the same sequence of each train.

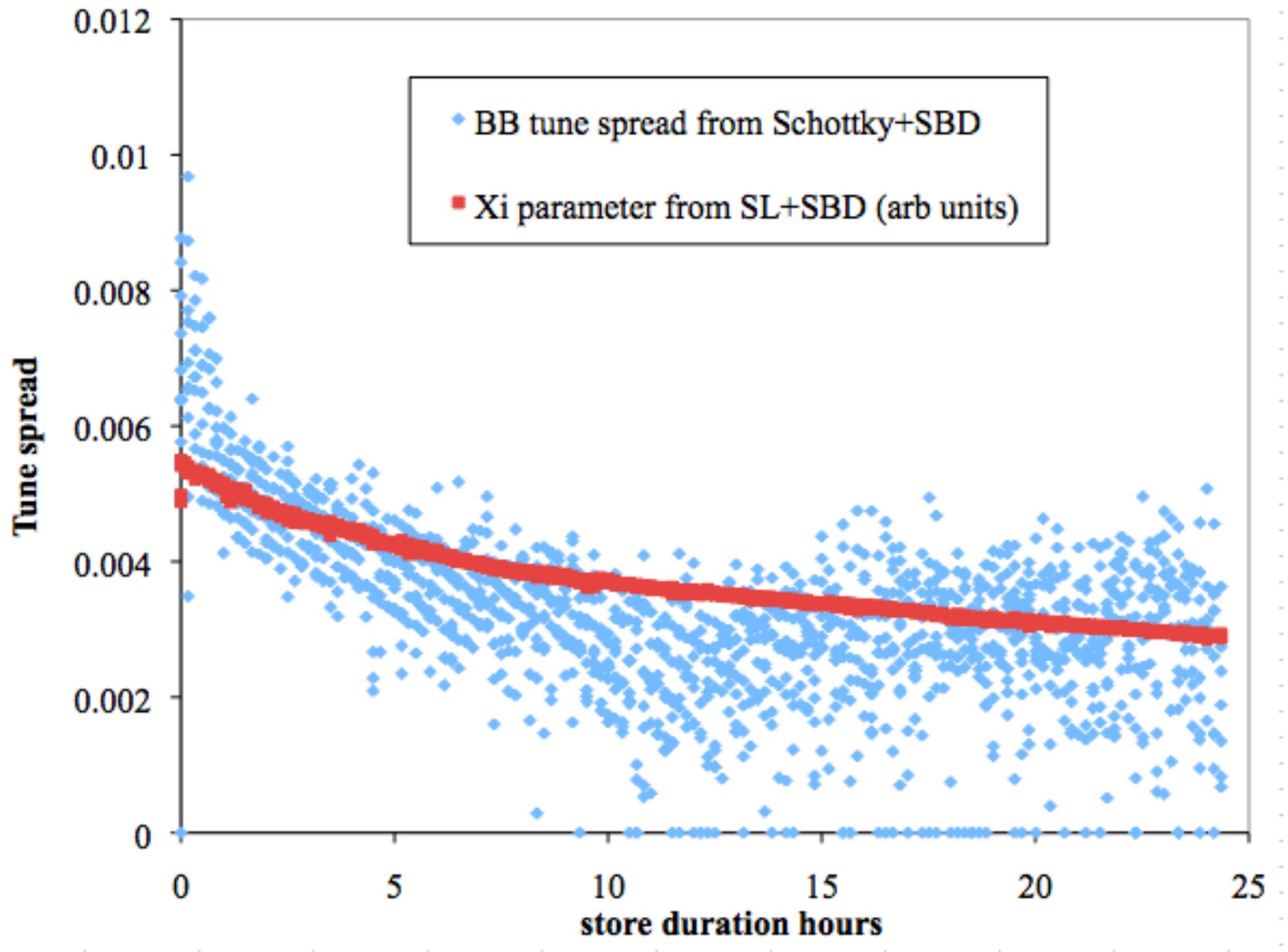

Figure 14. Difference, expressed in tune units, between frequency spread measured by Tevatron Schottky, and the frequency spread expected from beam momentum spread (as measured by Sample Bunch Display, SBD). The difference has the expected magnitude for the beam-beam spread, and scales as the beam-beam parameter $\xi$ (magenta line), as calculated from the transverse emittances (measured by synchrotron light, SL) and bunch length (from SBD).

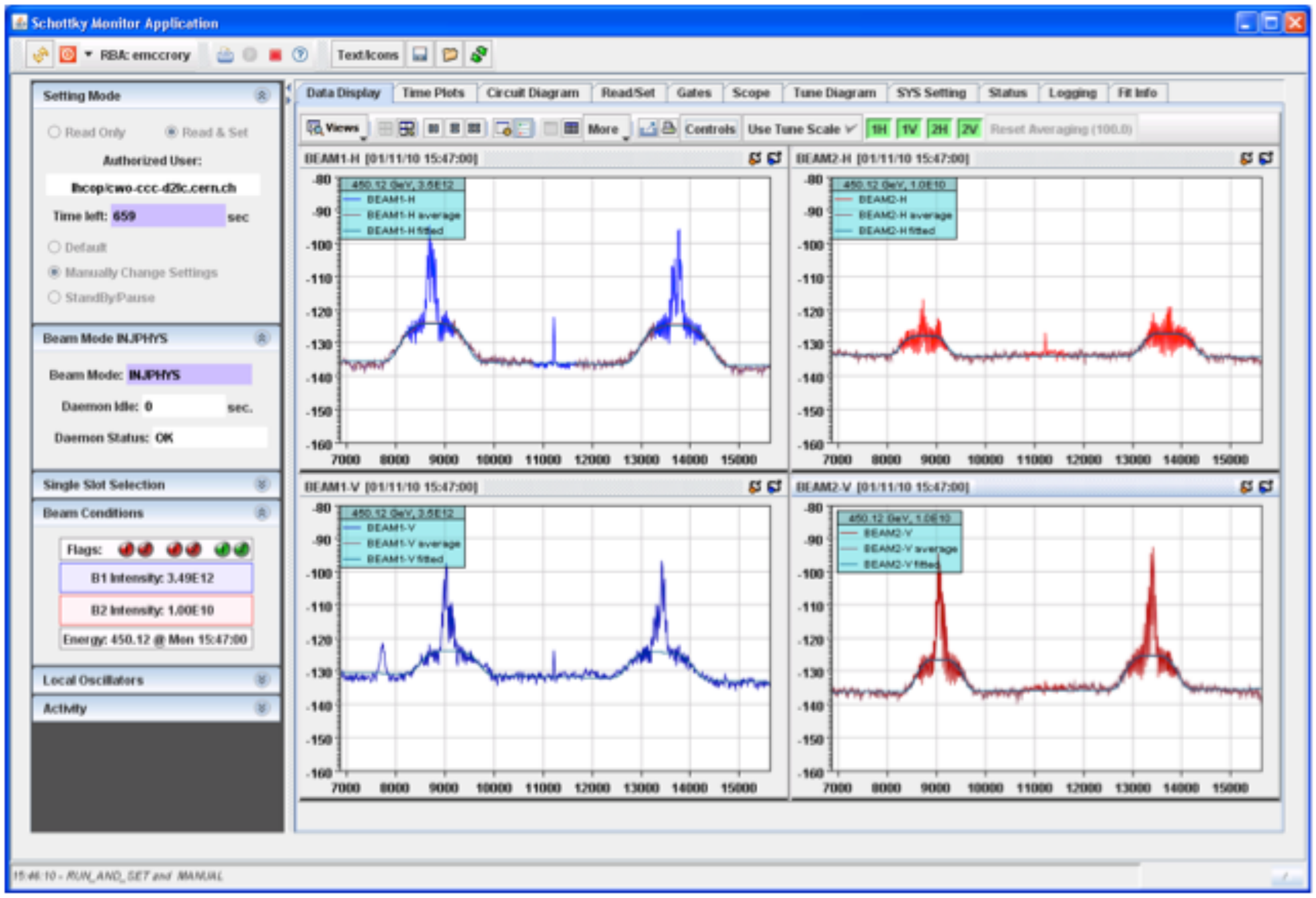

Figure 15. LHC curve fit data for protons at 450 GeV. Note significant coherence. The curve fit program eliminates the coherence from the fit. Center frequency is precisely at the half revolution harmonic and chosen via LO frequency settings. Beam 1 (blue) $3.49\times10^{12}$, Beam 2 (red) $1.4\times10^{12}$ Protons.

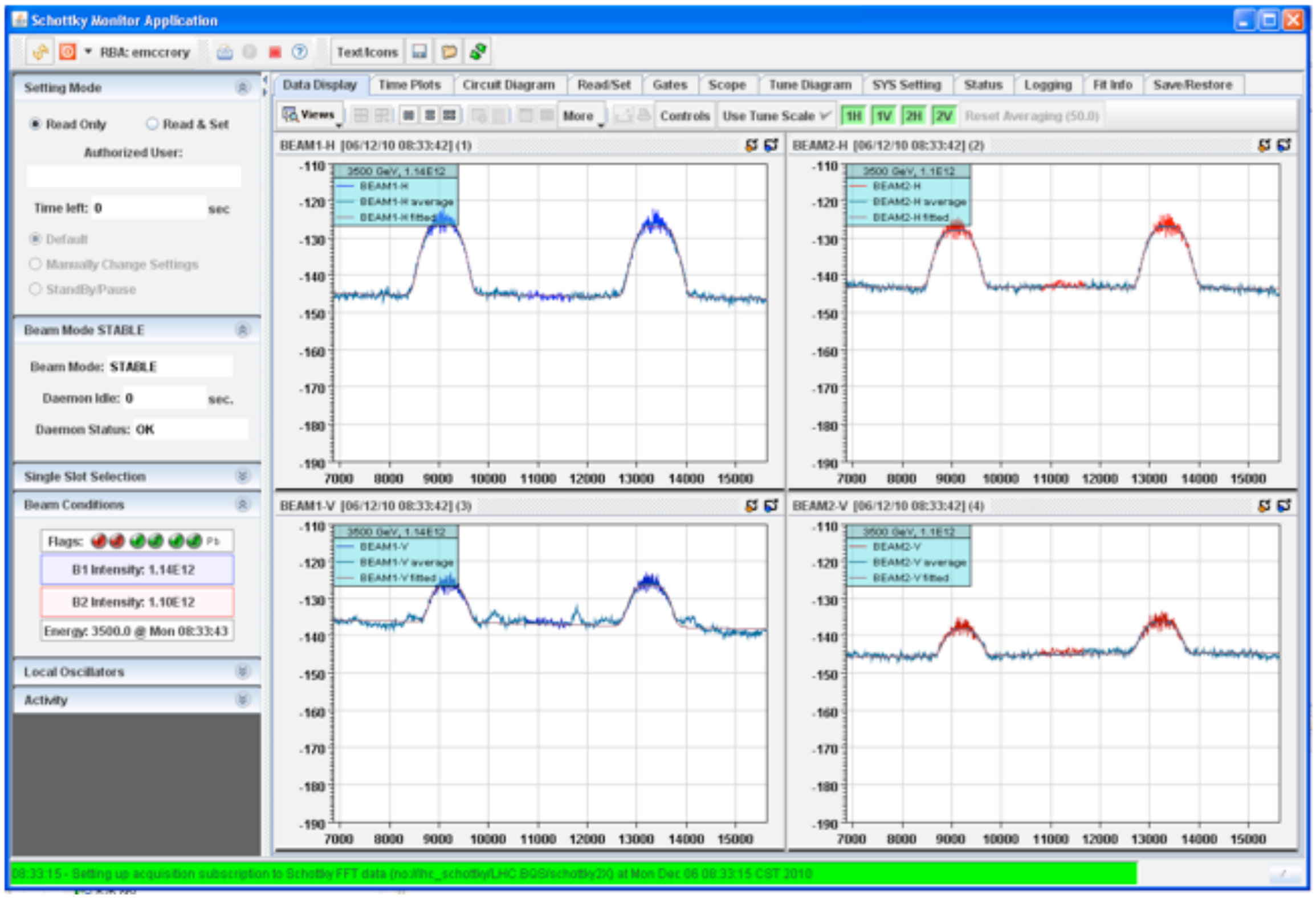

Figure 16. LHC Schottky curve fit data lead ions at 3500 GeV $1.1 \times 10^{12}$ per beam.

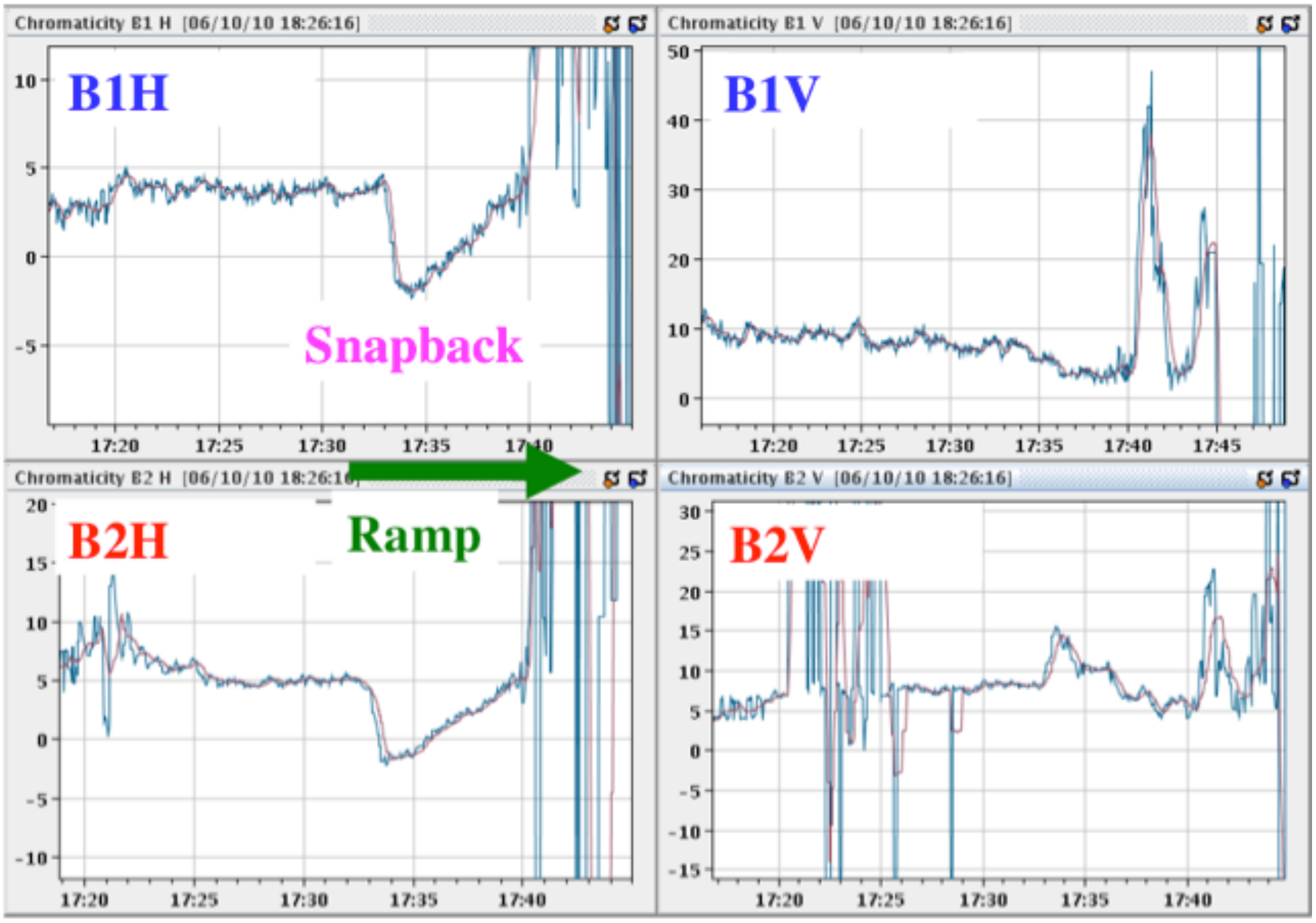

Figure 17. Control panel for chromaticity vs. time for protons from LHC Schottky application program.